# Divine Emanation As Cosmic Origin: Ibn Sînâ and His Critics

## Syamsuddin Arif*

Institut Studi Islam Darussalam (ISID) Gontor
Email: tagesauge@yahoo.com

## Abstract

The question of cosmic beginning has always attracted considerable attention from serious thinkers past and present. Among many contesting theories that have emerged, that of emanation was appropriated by Muslim philosophers like Ibn Sînâ in order to reconcile the Aristotelian doctrine of the eternity of matter with the teaching of al-Qur'ân on the One Creator-God. According to this theory, the universe, which comprises a multitude of entities, is generated from a transcendent Being, the One, that is unitary, through the medium of a hierarchy of immaterial substances. While the ultimate source is undiminished, the beings which are emanated are progressively less perfect as they are further removed from the first principle. The process is conceived as being atemporal and often compared to the efflux of light from a luminous body, or to water flowing from a spring. This metaphysical theory has enabled Ibn Sînâ to solve the vexed problem: given an eternally existing world and one eternally existing God, how can the two necessarily co-exist without having the perfect, simple unity of God destroyed by contact with the multiplicity of material things? The following essay delineates and evaluates both Ibn Sînâ's arguments as well as the counter-arguments of his critics.

Masalah permulaan kosmos selalu menarik perhatian para pemikir yang serius di masa lalu dan masa kini. Di antara berbagai teori yang muncul adalah teori emanasi yang telah dimodifikasi oleh filsuf Muslim seperti Ibn Sina untuk menyesuaikan doktrin Aristoteles tentang keabadian materi dengan ajaran al-Qur'an tentang Tuhan sebagai Pencipta Yang Esa. Menurut teori ini, alam, yang terdiri dari berbagai entitas, diturunkan dari Wujud yang transenden, Yang Satu,

* Pascasarjana Institut Studi Islam Darussalam (ISID) Gontor, Ponorogo, telp. (0352) 488220

yang Esa, melalui hirarki substansi material. Sementara sumbernya tetap dan tidak berkurang tingkatannya, wujud-wujud yang beremanasi secara progresif menjadi kurang kesempurnaannya ketika semakin jauhnya dari prinsip pertama. Proses ini dipahami sebagai tidak temporal dan seringkali dibandingkan dengan pancaran cahaya dari benda yang bercahaya, atau seperti air yang mengalir dari pancuran. Teori metafisika ini telah memungkinkan Ibn Sina menyelesaikan problem yang rumit yaitu: bagaimana dua wujud yang pasti yaitu wujud alam yang abadi dan wujud Tuhan yang juga abadi dapat berada secara bersamaan tanpa merusak kesempurnaan dan keesaan Tuhan yang berhubungan dengan keragaman benda-benda materi? Tulisan berikut ini membahas dan mengevaluasi argumentasi Ibn Sina dan juga argumentasi yang mengkritiknya.

## Preface

Few philosophers exercised as much and profound an influence on medieval Islamic, Jewish and Christian thought as Ibn Sînâ (d. 428/1037) or Avicenna as he was known in Latin.[1] He represents a development in philosophical thinking away from the apologetic concerns for harmonizing religion with philosophy towards an attempt to make philosophical sense of key religious doctrines. In this article I shall present his metaphysical cosmology and examine his views about the universe as an eternal, inevitable emanation or 'overflow' of the Divine One, rather than as something He created out of nothing. I shall discuss also the critical responses to Ibn Sînâ's views as put forward by subsequent thinkers, notably al-Ghazâlî (d. 555/1111) and Fakhr al-Din al-Râzî (d. 606/1210).

## The Problem of Beginning

Does the universe have a beginning? Is it created and originated or not? Although Ibn Sînâ's answer to these crucial questions turns out to be in the negative, the full account and detailed arguments by which he corroborates his theses are by no means simple and therefore deserve careful examination. Briefly stated, Ibn Sînâ's position on this issue is a kind of synthesis between two rival schemes. The standard doctrine held by the majority of Muslim theologians affirms (1) that the universe, by which is meant the physical world of matter, does have a 'beginning in time'—a definite moment in the past at which it was originated or created; (2) that its creator is one transcendent, eternal God; (3) that God's act of creation is voluntary in the sense that it is neither necessitated nor due to His eternal Essence; and (4) that God created it *not* from anything pre-existent whatsoever, but 'out of nothing' (*la min syay'*), which also means that its origination is preceded 'in time' by non-existence (*'adam*).

---

[1]On his biography and philosophical works, see W. Gohlman, *The Life of Ibn Sina* (Albany: SUNY Press, 1974); D. Gutas, *Avicenna and the Aristotelian Tradition* (Leiden: E.J. Brill, 1988); J. McGinnis, *Avicenna* (Oxford: Oxford University Press, 2010).

The other theory, though affirming God's existence, contends that the universe, or rather the constituent matter underlying it, is uncreated and that it has been there, eternally co-existing with God. This is the view subscribed by most philosophers (*falasifah*), who also deny God's active role in the affairs of the world and construe Him as nothing more than the First Cause, an automatically operating force that keeps the spheres in motion and thereby sustains the world. Ibn Sînâ's strategy was to adopt and appropriate the supposedly Aristotelian but actually Plotinian emanation scheme and fuse it with the Aristotelian metaphysics of self-sufficiency, causal necessity, and continuity of nature as well as with the Islamic monotheistic conception of the urgent contingency and immediate dependence of the world on God.

It is in a bid to reconcile those polarized opinions as well as in a move to preserve the Qur'anic conception of God as the One that Ibn Sînâ adopted and appropriated the theory of emanation in his cosmology, viewing the universe as the necessary outflow or emission from an eternal, necessitating cause, i.e. the 'necessarily existent' God, while at the same time maintaining the metaphysical distinction between essence and existence with respect to necessary and contingent beings. This theory warrants Ibn Sînâ to allow the emergence of multiple things in the universe from the One without infringing in any way the simple oneness of that One, who is the source of the being of all things that exist.

Thus, unlike the theologians who employ the terms *khalq* and *iḥdâts* for creation and origination respectively, Ibn Sînâ distinguishes the terms *sun'* (making), *takwîn* (forming or producing), and *ibdâ'* (inventing), from *îjâd* (bringing into existence). And he reserves the term *fi'l* (acting) to describe Divine Agency, that is, the manner in which God 'effects' the universe in the sense of causing it to exist, or to be precise, necessitating its existence and sustaining it. It is in this sense that Ibn Sînâ uses and understands the term "act", as distinguished from the ordinary usage. Thus he explains:

> When we say, "He is the agent of the universe (*fâ'il al-kull*)" we do not mean that He is the one who bestowed existence upon everything from scratch (*jadîdan*) which never before existed, as people commonly understand by it. For such an understanding presumes the following claims: either [1] this agent is called an agent insofar as existence emanates from it, or [2] insofar as existence does not come out of it, or [3] in both respects. Now if [it is an agent] insofar as existence issues from it, without taking into account the [previous] not-being of that existence, then [this is untenable because] the best agent should be the one from which existence emanates eternally. As for the second claim, it is obviously a contradiction. Finally, if it is an agent because it bestowed existence upon that which [previously] did not exist and at the same time does not give them existence, then it would avail nothing [that is, it was useless since it did nothing] when they did not yet exist. For non-existence does not require any cause, and even necessitates no cause. Indeed, its actual function consists in the fact that everything else receives its existence from it.[2]

Apart from *fi'l*, material creation (*takwîn*), and temporal origination (*iḥdâts*), there is, however, another plausible term which in his view could also convey his metaphysical notion of creation, namely the term *ibdâ'*.[3] But Ibn Sînâ immediately remarks that this should not be taken as the masses understand it. For in popular usage, the term *ibdâ'* simply means origination not from matter (*al-ikhtirâ' al-jadîd lâ 'an mâddah*), that is, creation out of nothing, whereas according to the the philosophers, *ibdâ'* signifies the eternal entification of that which in itself is non-entity (*idâmat ta'yîs mâ bi dhâtihi lays*), the process being eternal in the sense that the outcome or effect depends on and is necessitated by the essence of the First alone.

For Ibn Sînâ, that creation or creative process must be eternal follows from the fact that it depends on the essence of the First alone, its true Efficient cause, rather than on matter, instrument, idea, or any intermediary. The two terms thus differ in that whereas *fi'l* denotes the bestowal of existence and removal of non-existence at one time but not continuously, the term *ibdâ'* however means the same act (of giving and depriving existence) but with lasting effects.[4] Only in this sense can the term *ibdâ'* be substituted for the term *fi'l*. As can be seen, already in his terminology Ibn Sînâ indicates his preference for emanative creation over the well-received opinion that the universe was originated out of nothing.

---

[2] *Al-Mabda'*, 76.
[3] *Al-Isyârât: al-Ilâhiyyât*, 95.

## Existence as Divine Overflow

As far as Ibn Sînâ's doctrine of emanative creation is concerned, one could easily discern that the theory is grounded on a presupposed premise derived from Revelation and philosophy which states that God is a unique and absolutely simple Being.[5] Thus he maintains that the effect of God's creative activity has to be consistent with His 'unique' nature (that is, His absolute oneness), so that the effect too would have to be numerically one and substantially simple, and hence his famous formula: "from one thing only one thing could proceed (*al-wâhid min haytsu huwa wâhid innamâ yûjad 'anhu wâhid*)."[6]

However, since the universe is plainly composed and consists of innumerable things, it could hardly be supposed to have proceeded from God directly. Therefore, Ibn Sînâ contends, the only plausible explanation for the universe deriving its multiplicity from a single cause is to envisage a continuous series of individuals of various kinds proceeding from other causally prior entities, which serve as intermediaries between the First (*al-Awwal*) and the universe.[7] In Ibn Sînâ's view, the increasing scope and complexity of these intermediary causes and effects, ranging from the immaterial 'first intelligence' to the lowest of material things, would eventually account not only for the tremendous diversity of the world-system, but also for its causal origin and dependence upon God. Thus God is seen as the agent or efficient cause (*al-'illah al-fâ'iliyyah*) of the universe and the latter as the expression of His act.

---

[4] *Al-Mabda', 77.*

[5] The Qur'ân 112:1, *"Qul huwa Allâh ahad"*; id. 37:4, *"Inna ilâhakum lawâhid."*

[6] *Al-Syifâ': al-Ilâhiyyât*, 405; *al-Isyârât: al-Ilâhiyyât*, 216-7. On this wonderful formula see Mokdad A. Mensia, *Essai sur le principe «de L'un ne procède que de l'un» dans la philosophie islamique* (Thèse 3ème cycle; Paris, 1977) and Alain de Libera, "Ex uno non fit nisi unum. La lettre sur le principe de l'univers et les condamnations parisiennes de 1277," in *Historia philosophiae medii aevii.* [Festschrift für Kurt Flasch zu seinem 60. Geburtstag], ed. Burkhard Mojsisch and Olaf Pluta, 2 vols. (Amsterdam and Philadelphia, 1991), 1: 543-60.

[7] *Al-Syifâ': al-Ilâhiyyât,* 405.

Ibn Sînâ uses several terms to describe emanation, namely: *ṣudûr* (procession), *fayḍ* (overflow), and *luzûm* (necessary consequence).[8] His choice of these terms reflects at least two assumptions, namely, his view that (1) the actuality of every contingent being represents the existential plenitude and activity of that from which it emanates, and that (2) such actualization is necessarily outgoing and self-revealing in the sense that the act not only belongs to it but also extends outward from it.

Consequently, the procession of causes and effects will be continuous with its ultimate source in both a temporal and an ontological sense; temporally, it will be co-existent with God's creative activity, and ontologically, the causal series will remain inseparable from God simply because it is a necessary overflow of Himself.[9] This is part of the reason why Ibn Sînâ holds that the eternity of God's existence necessitates a co-eternal universe which is the collective embodiment of the emanation, and this is why for him the universe (*al-kull*)—that is, the totality of things constituting the physical world, though not identical with God, is somehow a projection out of Divine Plenitude.

According to Ibn Sînâ, it is from the Necessary Being, namely God, which is described as Pure Intelligence (*'aql mahd*) and the First Principle (*al-mabda' al-awwal*) that all other beings derive their existence, not directly but through intermediary (*bi wâsitah*). He insists, however, that we must not suppose that the universe comes into existence because God intended so (*'alâ sabîl qaṣd minhu*), for then He would act for something lower than Himself and introduce multiplicity (*takatstsur*) within His divine essence. Nor can it be the case that the universe comes into existence naturally by itself (*'alâ sabîl al-ṭab'*) in the sense that He is not aware of its genesis and does not mean it (*lâ bi ma'rifah wa lâ riḍâ minhu*).[10]

Indeed, God was and always is completely cognizant of both the universe (which is His effect) and the goodness emanating from Him—a fact which not only reflects His perfection but also manifests part of the necessary consequences of His majestic nature to which

---

every being yearns to return. It is by virtue of His act of self-reflection that the universe comes into existence (*ta'aqquluhu 'illah li'l-wujûd*) as a necessary consequence of His own Existence. Even so, Ibn Sînâ remarks, each of the issuing effects, including the universe is by no means identical (*mubâyinan*) with Him.[11]

Now, the First Principle is designated as Necessary Being (*wâjib al-wujûd*) in a double sense: not only does He exist necessarily, but He must act necessarily as well, and His act is an act of self-reflection. It is through His contemplation of His own essence (*ya'qil dhâtahu*) that the first effect (*al-ma'lûl al-awwal*), which is also said to be a pure, immaterial intelligence, necessarily proceeds. Since multiplicity (*kathrah*) is inconceivable in Him, the effect must be single (*wâḥid bi'l-'adad*), for as a rule, from one simple thing, only one can proceed.[12]

However, this formula breaks down in subsequent emanations. For as Ibn Sînâ tells us, within the first intelligence (*al-'aql al-awwal*) lies the germ of multiplicity, since its thought involves three acts of reflection, namely: (1) recognition of God's necessary existence, (2) consciousness of its own causally necessitated existence, and (3) awareness of its own existence as in itself only possible.[13] Consequently, the first act gives rise to another intelligence, the second act produces a celestial soul of the outmost sphere (*nafs al-falak al-aqṣâ*), whereas the third act generates the body (*jirm*) of this same sphere. Then the second intelligence, in a similar fashion, gives rise to a third intelligence, to the soul of the second sphere of the fixed stars, and to the body of that sphere. From the third intelligence there likewise emanates another triad, namely, a fourth intelligence, the soul of the third sphere, and the body of the third sphere.

This emanation of intelligences, we are told, goes on successively, each giving rise to successive triads and is halted only with the production of the sphere of the moon and the tenth or last intelligence, otherwise called the Agent Intellect (*al-'aql al-fa''âl*) from which our material world of generation and corruption originated. This Active Intelligence, instead of begetting the soul and body of a sphere, begets human souls and the four elements, i.e. water, air, fire, and earth.[14]

---

[11] *Al-Syifâ': al-Ilâhiyyât,* 403; *al-Mabda',* 75-6.
[12] *Al-Syifâ': al-Ilâhiyyât,* 404.
[13] *Al-Syifâ': al-Ilâhiyyât,* 405; *al-Najât,* 313; *al-Mabda',* 79.
[14] *Al-Najât,* 313-4.

As noted earlier, Ibn Sînâ conceives the universe as consisting of nine concentric spheres (*aflâk*) with their corresponding souls (*nufûs samâwiyyah*) and bodies (*ajrâm 'ulwiyyah*), in addition to the ten intelligences (*'uqûl*).[15] In ascending order of the spheres he places, like Ptolemy did, the moon, Mercury, Venus, sun, Mars, Jupiter, and Saturn—called the 'wandering stars' or planets (*al-kawâkib al-mutahayyirah*), whereas the Fixed Stars (*al-tsawâbit*) and another yet unnamed celestial body are said to be attached to the second and the first, outermost sphere respectively.[16]

Thus each planetary celestial body is believed to have only a single sphere (*falak*) or orb (*kurah*)[17] to which it is attached and by which it is carried around at various distances from the earth. In Ibn Sînian cosmic system, each intelligence, being the teleological cause in every emanative triad, becomes the target of desire (*syawq; 'isyq*) for the celestial soul within the triad, causing the eternal circular motion of the third component of the triad, the celestial body. And given the eternal motion of the celestial spheres, Ibn Sînâ thus postulates that the emanative process too must be eternal in the sense that God, the eternal efficient cause, ever in act, necessitates the existence of an eternal effect, the universe.[18]

One might curiously ask, however, why the process stops at the tenth, so-called Active Intelligence and does not go on *ad infinitum*. To this Ibn Sînâ replies: while it is true that the necessary procession of multiplicity of beings from one intelligence implies plurality of aspects (*ma'ânî*) in it, the reverse is not. That is to say, it would be wrong to assume that plurality of aspects always implies the necessary procession of multiplicity of beings. Nor is it true that every intelli-

---

[15] Ibid., 313-4.

[16] *Al-Syifâ: al-Riyâḍiyyât: 'Ilm al-Hay'ah,* ed. M. Ridâ Mudawwar and I. Ibrâhîm Ahmad (Cairo, 1980), 463. Cf. Ptolemy, *The Almagest,* trans. R. C. Taliaferro, in *the Great Books of the Western World,* vol. 16 (Chicago: Encyclopaedia Britannica Inc., 1952), 270 (bk. 9, chap. 1).

[17] These two terms are used by Ibn Sînâ indiscriminately, besides the equally common one: *dâ'irah*. But according to al-Bîrûnî (d. 1048), "*dâ'irah* and *falak* are two terms that denote the same thing and are interchangeable. However, sometimes *falak* refers to the globe (*kurah*), particularly when it is moveable (*mutaharrik*) and therefore *falak* does not apply to the motionless. It is called *falak* only on account of its similarity with the whorl of the rotating spindle ('alâ wajh al-tasybîh bi falakat al-mighzal al-dâ'ir)." See his *al-Qânûn al-Mas'ûdî* (Hyderabad: Osmania Oriental Publications Bureau, 1954), 54-5.

[18] *Al-Syifâ': al-Ilâhiyyât,* 407.

gence having the same kind of aspects will produce the same kind of effects.[19]

What Ibn Sînâ seemingly wishes to say is that the outcome depends on the nature and power of each emanative intelligence; and as intelligences succeed one another, their power decreases, and since the Active Intelligence stands low in the hierarchy its power is no longer sufficient to produce eternal beings like those emanated by the intelligences above it. Nevertheless, Ibn Sînâ ascribes to the Active Intelligence a set of functions that lend his scheme a balance missing in that of al-Fârâbî, who assigns the Active Intelligence functions related solely to the actualization of the human mind.[20] By contrast, in Ibn Sînâ's scheme, the Active Intelligence, being the emanative cause of matter of our sublunar world (*'aql al-âlam al-ardî*),[21] is not only responsible for bestowing the earthly beings their natural 'forms' (i.e. their souls) but also in charge of (*yudabbiru*) the souls of humans, animals and plants.[22]

Furthermore, the Active Intelligence is also described by Ibn Sînâ as the cause of the actualization of human minds (*al-jawhar al-mukmil li anfus al-nâs*)[23] as well as the source of their intuitive knowledge.[24] No wonder then the Active Intelligence is often called the Giver of Forms (*wâhib al-suwar*)[25] and sometimes also identified as the Archangel Gabriel (*rûh al-quds*) or the Angel of Revelation (*al-rûh al-amîn*).[26]

## Criticism and Response

Ibn Sînâ's emanation scheme has stirred up debates and evoked polemical reaction. I shall review some of the arguments advanced by its prominent critics, notably al-Ghazâlî (d. 1111) and Fakhr al-Dîn al-Râzî (d. 1210). Al-Ghazâlî raises five objections (*i'tirâḍât*) to

---

[19] *Al-Syifâ': al-Ilâhiyyât*, 407; *al-Mabda'*, 80.
[20] See Al-Fârâbî, *al-Madînah al-Fadîlah*, ed. and trans. R. Walzer, *Al-Fârâbî on the Perfect State* (Oxford: Clarendon Press, 1985), 100-105.
[21] *Al-Najât*, 310.
[22] *Al-Najât*, 314; *al-Shifâ': al-Ilâhiyyât*, 410; *al-Mabda'*, 80.
[23] *Al-Syifâ': al-Ilâhiyyât*, 388.
[24] *Al-Syifâ': al-Tabî'iyyât: al-Nafs*, 208.
[25] *Al-Syifâ': al-Ilâhiyyât*, 413; *al-Syifâ': al-Tabî'iyyât: al-Nafs*, 218
[26] See Al-Fârâbî, *Kitâb al-Siyâsah al-Madaniyyah*, ed. with intro. and notes by Fawzi M. Najjâr, 2nd imp. (Beirut: Dar el-Machreq, 1993), 32.

Ibn Sînâ's theory of emanation.[27] First, he questions whether the 'being-possible' of the first effect, whose existence is said to be possible, is identical with its existence (*'ayn wujûdihi*) or not. If identical, then there is no plurality; but if different, then the being-necessary-in-itself of God's existence too must be other than His existence—a logical conclusion Ibn Sînâ would not allow because it implies plurality in the One.

This objection, however, might just as well be dismissed since al-Ghazâlî has, in the first place, seen no harm in affirming the presence of multiplicity in God, in accordance with the Ash'arite doctrine of God's various Names and Attributes.[28] Moreover, this criticism in fact stems from his outright rejection of the most fundamental thesis in Ibn Sînâ's metaphysics, that God's existence is necessary in itself. For al-Ghazâlî, to affirm the existence of God and deny the necessity of such an existence at the same time does not at all involve contradiction because God, being transcendent and unknowable, is beyond such human-invented concept.[29]

Both the second and third objections are likewise theological and even begging the question, arguing that God's knowledge involves the idea of multiplicity and so does His thought, which is precisely the point at issue. The fourth charge contends that the first effect, being a pure intelligence, is insufficient to produce something composed of form (soul) and matter (body), a particular size, axis, etc., like a celestial body. Finally al-Ghazâlî claims that he finds no convincing arguments offered by Ibn Sînâ that prove his assertion that emanation is a necessary process.

---

[27] See al-Ghazâlî, *Tahâfut*, 100-9. Cf. Michael E. Marmura, "The Conflict over the World's Pre-eternity in the *Tahâfut*s of al-Ghazâlî and Ibn Rushd," (Ph.D diss., University of Michigan, 1959), 20-4. Other critics include al-Shahrastânî, *Musâra'at al-Falâsifah*, ed. S. M. Mukhtâr (Cairo, 1976), 59-60 and 86-88, as well as Ibn Taymiyyah, *Minhâj al-Sunnah al-Nabawiyyah fî Naqd Kalâm al-Syî'ah wa al-Qadariyyah*, 4 vols. (Cairo: al-Matba'ah al-Amîriyyah, 1321 A.H.), 1: 89 and 1: 94-6.

[28] The Ash'arites maintain that God's Attributes exist *in* Him as eternal, separate immaterial entities, in contrast to the Mu'tazilites who hold that the doctrine of *tawhîd* necessitates that no entity exists in His Essence and there is a kind of identity between God and His Attributes. See 'Abd al-Qâhir al-Baghdâdî (d. 429 AH/ 1037 CE), *Kitâb Usûl al-Dîn*, 1st imp. (Istanbul: Matba'at al-Dawlah, 1928), 90 and 109, and al-Shahrastânî, *Nihâyat al-Iqdâm*, 204 in which he argues against the philosophers that even the notion of Necessary Being admits of duality, i.e. necessity and existence, so that if they (the *falâsifah*) can accept it, why then not the Attributes? For further discussion, see Michel Allard, *Le problème des attributs divins dans la doctrine d'al-Aš'arî et de ses premiers grands disciples* (Beirut: Imprimerie Catholique, 1965).

[29] See *Tahâfut*, 143-4; S. Dunyâ's ed.: 173-5 and 181.

For al-Ghazâlî, the whole account is simply absurd. Again, this objection rests on al-Ghazâlî's conception of God as a voluntary agent who can create plurality and diversity as He wishes and wills (*yakhluq al-mukhtalifât wa al-mutajânisât kamâ yurîd wa 'alâ mâ yurîd*).[30] But what Ibn Sînâ seeks, in contrast with al-Ghazâlî's appeal to Revelation, is a properly rational explanation that would fit well into his grand metaphysical system. Seen in this perspective, any criticism leveled against Ibn Sînâ's theory would count only if, the basic premises having been admitted, it succeeds to expose the internal logical inconsistencies that would bring down his system.

Interesting to note in this regard is Ibn Rushd's observation of a fundamental error committed by emanationist philosophers like al-Fârâbî and Ibn Sînâ. According to the Andalusian philosopher, by affirming the *ex uno non fit nisi unum* principle and then assuming multiplicity in the first entity that proceeds, the philosophers are forced to regard this multiplicity as uncaused,[31] a consequence which merely shows inconsistency in their use of the principle.

In order to resolve this apparent contradiction, however, one needs only to recall the fact that Ibn Sînâ, true to his strict monism, cannot allow more than one effect to proceed from the One precisely because His active intellection, reflection or contemplation (*ta'aqqul*) has been and is focussed purely and *only* on Himself, for in the beginning "only He and nothing else was,"[32] so that *only* one single effect could emanate. The same rule would have definitely applied if we encounter a similar situation in the next process, which is not the case. For the first intelligence, as Ibn Sînâ put it, is naturally conscious of its own self, cognizant of its Cause, and aware of the fact that its existence, considered in itself, is merely possible. It is these seeds of plurality existing in the first and subsequent intelligences which nullify the monistic principle when it comes to the nine succeeding emanations.

---

[30] *Tahâfut,* 109.

[31] *Tahâfut al-Tahâfut*, 249-50.

[32] A well-known tradition: *"kâna Allâh wa lam yakun shay' ma'ahu,"* related by al-Bukhârî, *al-Jâmi' al-Sahîh* (Beirut: Dâr al-Fikr, n.d.), 4: 129. Cf. *"Kâna Allâh wa lâ shay' ma'ahu,"* reported by al-Zabîdî, *Ithâf al-Sâdah al-Muttaqîn*, 2: 105 and *"kâna Allâh wa lam yakun shay' ghayruhu,"* narrated by al-Bayhaqî, *al-Sunan al-Kubrâ*, 9: 3 and al-Tabarânî, *al-Mu'jam al-Kabîr,* 18: 205.

Turning now to al-Râzî's objections (*syukûk*), we shall summarize and consider only the following point.[33] Generally, he complains of the ambiguity in Ibn Sînâ's statement about the emergence of plurality from the first effect; is it due to the possibility of its reflection of its own existence (*imkân ta'aqquli wujûdih*) or, as noted earlier, is it because of its very reflection of its own possibility (*ta'aqqul imkâni nafsih*), etc? Al-Râzî indeed refutes both possible readings.[34] Specifically, he questions whether in Ibn Sînâ's scheme a second intelligence and a celestial body emanated from the first intelligence because of the latter's being-possible in itself, or because of its being-necessary from another and its knowledge of its Cause. Both alternatives, he contends, are untenable. For neither possibility, nor necessity-by-virtue-of-another, nor existence can serve as a cause.[35] Al-Râzî adduces several logical proofs in support of his view that they cannot be causes. In the case of possibility, for instance, he argues that whatever is non-existent in the external world (e.g. a possible entity) cannot be the cause of something that does exist externally (e.g. a celestial body).[36]

Having proved that possibility is not an existential entity (*amr wujûdî*) and therefore cannot be the cause of anything existent,[37] al-Râzî then takes up each of the other attributes of the first intelligence, namely its possible existence, its necessity, its self-knowledge and its knowledge of its Cause, showing that none of them is sufficient to serve as a cause.[38] It is clear that most if not all arguments put forth by al-Râzî purport to overthrow the fundamental emanationist thesis that from a one thing only one can proceed.[39]

A sophisticated reply to al-Râzî's criticisms has come from Nasîr al-Dîn al-Tûsî (d. 1273 CE), who carefully explains two major problems in Ibn Sînâ's theory, that is, the issue of multiplicity in the first intelligence, and the question of exactly which aspects within the first intelligence are the causes of which effects.[40] The first intelligence, says al-û,sî, has a total of six aspects (*haytsiyyât*), of which two are its constituent parts (*muqawwimât*) namely its existence and quiddity, and the other four are its concomitants (*lawâzim*), which include its possibility in itself, its necessity through its Principle, its self-knowledge, and its knowledge of its Principle.[41] This is so because on Ibn Sînâ's account it is impossible for the first intelligence, which is a caused entity, to be composed of various things (*muqawwaman min mukhtalifât*).[42]

---

[33] For more discussion, see Nicholas Heer, "Al-Râzî and al-Tûsî on Ibn Sînâ's Theory of Emanation," in *Neoplatonism and Islamic Thought,* ed. Parvez Morewedge (Albany: SUNY Press, 1992), 115-8.

[34] Fakhr al-Dîn al-Râzî, *al-Mabâhits al-Masyriqiyyah,* 2 vols. (Qumm: Matba'at Amîr, 1991), 2: 503.

[35] See his commentary in *Kitâb Syarhay al-Isyârât,* printed on the margins, 2 vols. (Cairo: n.p., 1907), 2: 48 lines 20-3; cf. ibid., 2: 49 line 12ff.

[36] Ibid., 2: 49. Cf. Fakhr al-Dîn al-Râzî, *Lubâb al-Isyârât,* printed as in Ibn Sînâ, *al-Tanbîhât wa al-Isyârat* [sic!], ed. Mahmûd Shahâbî (Tehran: Tehran University Press, n.d.), 267-8.

[37] *Syarhay al-Isyârât,* 2: 48 lines 23-36.

[38] *Mabâhits,* 2: 503-5.

[39] *Lubâb,* 268.

But more importantly, al-Tûsî does not equate the first effect with the first intelligence. For him, the first effect to emanate from God was existence (*wujûd*), which he construes as merely an aspect or, to be precise, one of the two constituents of the first intelligence.[43] This is because, according to al-Tûsî, the term *al-ma'lûl al-awwal* is used equivocally, sometimes it designates the simple existence and at other times it refers to the composite first intelligence.[44] Coming to the second problem, he asserts that quiddity and possibility, which are considered to be non-existential (*'adamiyyayn*) in themselves and existential only *ab alio*, and which represent the state of the first intelligence in its potentiality, are responsible for the matter (body) of the celestial sphere. In contrast, existence and self-knowledge, which represent its state in actuality, are responsible for the form (soul) of the sphere.[45] It is the last two aspects, namely, necessity and knowledge of the First Principle, which represent the state of the first intelligence insofar as it is derived from God, that are responsible for the emanation of another intelligence. Al-Tûsî concedes that none of these aspects is existential entity and therefore cannot be independent causes in themselves. But they do serve, he insists, as conditions (*syurût*) and modes (*ḥaytsiyyât*) through which the true Efficient Cause acts and creates.[46]

---

[40] Cf. N. Heer, "Al-Râzî and al-Tûsî on Ibn Sînâ's Theory of Emanation," in *Neoplatonism and Islamic Thought,* 119-23.

[41] See his commentary in *al-Isyârât: al-Ilâhiyyât,* 219 and 221.

[42] Comment in *al-Isyârât: al-Ilâhiyyât,* 223.

[43] Comment in *al-Isyârât: al-Ilâhiyyât,* 218.

[44] Comment in *al-Isyârât: al-Ilâhiyyât,* 226.

[45] Comment in *al-Isyârât: al-Ilâhiyyât,* 223-4.

[46] Comment in *al-Isyârât: al-Ilâhiyyât,* 225.

The emanation theory has also led some people to charge Ibn Sînâ with pantheism. They argue that to regard the universe and everything there as an emanation from the One is to blur the distinction between the Creator and creatures. This criticism, however, happily ignores the clear statement made by Ibn Sînâ that "He is the Existent from which each and every existence emanates; His Existence is Essential and distinct (*mubâyin*) from every other existence."[47]

True, there is a big difference between creating from nothing and producing from one's thought. In the latter case, as Morewedge points out, a resemblance is implied between the source and its outcome.[48] But as a matter of fact, Ibn Sînâ does postulate a Being utterly transcendent with respect to all other beings, in spite of his adherence to emanationism when it comes to the question of creation. In his scheme, as indicated earlier, the gulf separating the transcendent God and eternally emanated hierarchy of beings is bridged by the First Intelligence, which in one text is identified with the first Archangel-Cherub.[49] Another guarantee against any danger of pantheistic interpretation is to be found in Ibn Sînâ's famous if not enigmatic doctrine of essence and existence.[50]

## Conclusion

To sum up, the theory of emanation was meant to supplement the meagre and Islamically unacceptable view formulated by Aristotle to whom there was no passage from God, the One, to the world, the many. The theory was apparently intended less as an account of the origin of the universe than a description, in temporal imagery, of the eternal relation of the world to God. For in Ibn Sînâ's view there is no absolute beginning of a finite being here since, according to him, a beginning refers not to just one *now* of time, but to every time and age.[51] It is impossible, he says, that a thing begins to be after it was not, since [prime] matter would precede it from which it would begin to be.[52] And like al-Fârâbî, he was trying to reconcile the Aristotelian doctrine of the eternity of matter with the teaching of al-Qur'ân on the One Creator-God.

---

[47] *Al-Mabda'*, 76. Cf. *Al-Syifâ': al-Ilâhiyyât*, 403.

[48] P. Morewedge, *The Metaphysica of Avicenna (ibn Sînâ). A Critical Translation-Commentary and Analysis of the Fundamental Arguments in Avicenna's Metaphysica* in the *Dânish Nâma-i 'alâ'î* (The Book of Scientific Knowledge), Persian Heritage Series, no. 13 (New York: Columbia University Press, 1973), 272.

[49] Henry Corbin, *Avicenna and the Visionary Recital* (Dallas: Springs Publications, 1980), 58.

[50] This subject has been discussed by Fazlur Rahman, "Essence and Existence in Avicenna," *Medieval and Renaissance Studies* 4 (1958): 1-16; id., "Essence and Existence in Avicenna: Myth and Reality," *Hamdard Islamicus* 4 (1981): 3-14; and P. Morewedge, "Philosophical Analysis and Ibn Sînâ's Essence-Existence Distinction," *Journal of the American Oriental Society (JAOS)* 92 (1972): 425-35.

Indeed, Ibn Sînâ's emanation theory represents an attempt to solve this vexed problem: given an eternally existing world and one eternally existing God, how can the two necessarily co-exist without having the perfect, simple unity of God destroyed by contact with the multiplicity of material things? Ibn Sînâ's answer was to interpose many levels of spiritual substances, namely, the intelligences, between God and matter as a shield to safeguard and maintain the divine Oneness from multiplicity. In other words, although God remains in Himself and high above transcending the created world, there are, nevertheless, intermediary links between the absolute eternity and necessity of God and the world of downright contingency. Thus by relating the spiritual intelligences to God as the necessarily acting Source of their beings, Ibn Sînâ is able to account both for the necessity of their being and for their indebtedness to God as their Efficient Cause.[]

---

[51] See *al-Mabda',* 45 and *al-Najât,* 176.

[52] *Al-Najât,* 224. Although the term used here is simply *mâddah,* one can understand from the context that Ibn Sînâ is referring to the prime matter (*al-hayûlâ*).